\documentclass[11pt]{article}
\usepackage{geometry}                % See geometry.pdf to learn the layout options. There are lots.
\usepackage{graphicx}
\usepackage{amssymb}
\usepackage{epstopdf}
\usepackage{hyperref}
\usepackage{listings}
\usepackage{color} %red, green, blue, yellow, cyan, magenta, black, white
\definecolor{mygreen}{RGB}{28,172,0} % color values Red, Green, Blue
\definecolor{mylilas}{RGB}{170,55,241}
\title{\Large\bf Responsive Graphical User Interface (ReGUI) \\ and Its Implementation in MATLAB }
\author{\textbf{Matej Mikulszky} \\ 
\textbf{Jana Pocsova}\\
\textbf{Andrea Mojzisova}\\
\textbf{Igor Podlubny\footnote{Corresponding author}}\\[1.5ex]
BERG Faculty, Technical University of Kosice, \\
B. Nemcovej 3, 04200 Kosice, Slovakia\\[1ex]
matej.mikulszky@student.tuke.sk, \\
jana.pocsova@tuke.sk\\
andrea.mojzisova@tuke.sk\\
igor.podlubny@tuke.sk
}
\begin{document}
\maketitle

\begin{abstract}
\noindent
In this paper we introduce the responsive graphical user interface 
(ReGUI) approach 
to creating applications, and demonstrate how this approach can be implemented in MATLAB. 
The same general technique can be used in other programming languages.
\end{abstract}

\section{Introduction}

The term \emph{Responsive Web Design}, or RWD, was introduced by Ethan Marcotte 
in his online article in 2010 \cite{Marcotte-RWD-blog}.
It conquered the world of web  immediately, and has since been
widely used in design of web pages and web-based applications. 

% Hans Selye, page 6, 
% it is not to see something first, but to establish solid connections 
% between the previously known and the hitherto unknown 
% that constitutes the essence of scientific discovery.

In fact, this approach was, at least in parts and in some examples of web design, 
known earlier, but was not referred to as RWD. 
However, as Hans Selye wrote~\cite{Selye-Stress-Book}, the discovery  
\emph{``is not to see something first, but to establish solid connections 
between the previously known and the hitherto unknown''}. 
From this point of view, Ethan Marcotte is generally recognized
as the one who formulated the principles of the responsive web design and provided a general way to its
implementation on the web.

The main idea of the RWD is the division of the content of a web page into blocks, 
which can be arranged in several logical ways in order to optimize the user experience. 
This, of course, reminds us the modular approach to typography \cite{West-book}; 
however, instead of a fixed-sized sheet of paper, a resizable browser window 
is the media for delivering the content. This different media prompts the modular approach 
to be enhanced by adding ``fluidity'' to it. 

In this paper we take the idea of the ``responsive design'' further, apply it 
to the design of GUIs for stand-alone applications, and demonstrate 
how this can be done in MATLAB. The same technique can be used 
in other programming languages. 

% MARTIN:
% It should be mentioned, that while describing the approach presented in this paper, 
% one can encounter terms they are familiar with, but those terms have different meaning 
% in this context.
It should be mentioned that one can encounter some terms that sound similarly 
to the approach presented below in this paper, but they have different meaning.
For instance, the term ``responsive user interface'' can in some contexts denote situations when the main process
of an application is busy or not responding (frozen), while the GUI is still able 
to accept (to respond to) user's inputs in the form of mouse clicks or keystrokes. 
Another such example is the term ``adaptive user interface''. This term often describes situations when the user interface
adapts to the role of a user, such as switching between basic and advanced modes, 
or switching between user roles such as ``teacher/student'', ``doctor/patient'', ``admin/user'', etc.

\section{Responsive GUI for applications}

Recall that the responsive web design is based on the idea that a web page must
respond by itself to the user's behavior and the constraints of the media environment imposed by its screen size, pixel density, platform and orientation. 

Hereby we extend this approach to designing the graphical user interface (GUI) 
of stand-alone applications, and arrive at what we call the 
\emph{responsive graphical user interface} --
ReGUI\footnote{Finding a suitable -- and available -- abbreviation is always a problem\ldots 
We ended up with \emph{ReGUI};  we found that in Catalan the word \emph{regui} 
 is one of the forms of the verb \emph{regar} (`to water'), 
which is well related to `fluidity' of the responsive graphical user interface.}.

Before continuing, it is worth mentioning, that some elements of the responsive approach 
can already be observed in users' interaction with computer programs.

First, many contemporary applications have the form of web applications.  
Their interface is a web page, so it is natural that 
for creating the user interface the RWD approach is used.
Our focus, however, is on stand-alone applications.

Second, calculators  for smartphones  provide another example 
of responsive design -- in portrait mode, a user sees a simple primary-school-like calculator, 
but rotating the smartphone to landscape transforms it into a scientific calculator 
with a different set of buttons and functions. Some other programs, like mobile web browsers, 
also provide similar functionality. 

Third, touch keyboards in smartphones and tablets are  also ``responsive'',  
since switching to different languages changes the layout and the number of  ``touch keys'', 
additional keys appear specifically for particular applications, and so forth. 
This is somewhat similar to the behavior of aforementioned calculators for smartphones,
because one whole set or subset of keys is shown while another is hidden. 

Also, in many desktop applications, like current word processors, 
spreadsheets, etc.,  separate document windows  are opened or created in positions based 
on the number and physical dimensions of the computer screens available.

We define the basic requirements to the ReGUI in applications as follows:
\begin{itemize}
\item [(R1)]
independence of the physical dimensions and resolution of the user's device screen;
\item [(R2)]
ability to zoom in and out, preserving all proportions in the ReGUI elements, including fonts;
\item [(R3)]
responsiveness of the layout of the logical blocks and the elements of the ReGUI based on the user's needs 
that are indirectly indicated by  his/her behavior (resizing the application window, moving it, etc.).
\end{itemize}

Since we use MATLAB as a tool for scientific computations and applications 
in modelling and control of processes, we elaborated several examples 
of creating applications with different levels of ReGUI. Our choice of MATLAB 
is also based on the high readability of the MATLAB code. 

The re-engineering of the ``Matrix Poker''  game~\cite{Matrix-Poker}, 
developed by the last author in 2003 in a fixed-size form with physical units of pixels, 
was done in October 2016 by consistently transforming  all dimensions and font sizes 
to relative units (in MATLAB terminology, \emph{normalized} units). 
The width-to-height aspect ratio is preserved, except for  the situation, 
where the game window after resizing occupies the whole screen. 

Creation of a simple zoomable application from scratch is easier, if we think in terms of relative (normalized) units from the very beginning. This is illustrated by the application Anaglyph3D~\cite{Anaglyph3D} for making fine adjustments (micro-shifts and micro-rotations) of the left and right images of stereo photos 
that were taken separately using a single camera without a special dedicated tripod. 

Both of these applications, Matrix Poker and Anaglyph3D,   
satisfy the first two requirements to the application's ReGUI, i.e. requirements (R1) and (R2), 
but they do not contain enough logical blocks of the GUI elements to illustrate the core 
idea of ReGUI represented by the requirement (R3). 
How can one satisfy the requirement (R3) is shown in detail in the following example.

\section{Creating a ReGUI application in MATLAB}

The application \texttt{TeachLCGE}~\cite{TeachLCGE}, that we describe below, 
has been developed for supporting teaching maxima and minima of functions of two variables. 
The source code is available at MATLAB Central File Exchange~\cite{TeachLCGE}, 
and the video demonstration of its ReGUI functionality is available on YouTube~\cite{TeachLCGE-video}.

The logical blocks that we consider in the user interface are:

\begin{itemize}
\item
\textsc{Function}, with an editable text field for entering the formula for $f(x,y)$ 
and three buttons named  \texttt{Show Example},  \texttt{Calculate},  \texttt{Clear the form};
\item
\textsc{Stationary points};
\item
\textsc{Function values};
\item
\textsc{Local extrema};
\item
\textsc{Plots}, with four subplots; the subplots can be arranged as 2x2 or 1x4;
\item
\textsc{Show plots in separate windows}, with four buttons -- each button for a particular plot;
\item
a dedicated button \textsc{Back to main menu}, in which the user selects the type of the problem and the language (currently, English or Slovak).
\end{itemize}

These logical blocks are laid out in a different manner  based on the width-to-height aspect ratio $R$
of the resizable window of our application. In this example, we set up two breakpoints 
for changing the ReGUI layout, namely 0.75 and 1.5. When the aspect ratio $R$ 
is between 0.75 and 1.5 (inclusive), the application window looks more or less similar 
to most computer screens with the old classical aspect ratio 4:3;
when $R > 1.5$, then it is similar to wide screens in landscape orientation; 
and when $R < 0.75$, then it looks like a wide screen rotated 90 degrees to the portrait orientation. 

It is important that our application window is fully resizable, so a user can change the width
and the height of the application window independently, and as the result the layout 
of the all logical blocks and elements of the application is adapted to a new aspect ratio 
of the application window. 

Let us very briefly outline the implementation of ReGUI. 
The first line numbers in the code fragments below correspond to the appropriate lines
in our MATLAB application \texttt{TeachLCGE}~\cite{TeachLCGE}. 

Similar to the case of responsive web design, we start with defining the layout breakpoints 
(and, of course, mocking up the corresponding variants of the layout):

\begin{lstlisting}[frame=single, xleftmargin=1cm, xrightmargin=1cm,firstnumber=1043]
% Layout aspect ratio breakpoints
    breakPoint = [0.75 1.5];
\end{lstlisting}

\noindent
The application window aspect ratio is computed as usual:
\begin{lstlisting}[frame=single, xleftmargin=1cm, xrightmargin=1cm,firstnumber=1043]
% Aspect Ratio
    appWinPosition = get(appWindow,'Position');
    appWinWidth  = appWinPosition(3);
    appWinHeight = appWinPosition(4);
    aspectRatio  = appWinWidth/appWinHeight;
\end{lstlisting}

\noindent
In the next step, the data for  the updated ReGUI layout is prepared (computed) based on the 
currently updated value of the aspect ratio:

\begin{lstlisting}[frame=single, xleftmargin=1cm, xrightmargin=1cm,firstnumber=1046]
 %  ReGUI section
if (aspectRatio >= breakPoint(1)) && (aspectRatio <= breakPoint(2))
    % Classic interface values
    panel0 = [0.01 0.75 0.38 0.175];
    titlePosition   = [0.25  0.94  0.5  0.04];
    grsurfPosition  = [0.01  0.935 0.12 0.04];
    %  < ------ and so forth ------ >
end

if (aspectRatio < breakPoint(1))
    % Tall interface values
    panel0 = [0.01 0.765 0.98 0.16];
    titlePosition   = [0.05  0.96   0.9   0.04];
    grsurfPosition  = [0.01  0.925  0.25  0.04];
    whitePosition   = [0.25  0.93   0.33  0.04];
    %  < ------ and so forth ------ >
end

if (aspectRatio > breakPoint(2))
    % Wide interface values
    panel0 = [0.01 0.65 0.38 0.27];
    titlePosition   = [0.05 0.93  0.9  0.06];
    grsurfPosition  = [0.01 0.92  0.25 0.06];
    whitePosition   = [0.10 0.935 0.12 0.05];
    %  < ------ and so forth ------ >
end
 \end{lstlisting}
 
 \noindent
 And finally the positions and/or the appearance of the ReGUI interface elements 
 are updated using the computed data:
    
\begin{lstlisting}[frame=single, xleftmargin=1cm, xrightmargin=1cm,firstnumber=1095]
if (Updater1 == 1)
    % Update values of properties of ReGUI elements
    set(frame0,'Position',titlePosition);
    set(frame1,'Position',examplePosition);
    set(frame2,'Position',clearPosition);
    %  < ------ and so forth ------ >
end
\end{lstlisting}

\noindent
The main steps that are described above are done by the function  \texttt{regui}, 
which is triggered by the \texttt{ResizeFcn} property of the main application window:
  
\begin{lstlisting}[frame=single, xleftmargin=1cm, xrightmargin=1cm,firstnumber=247]
% Main application window
appWindow = figure('Units','norm', ...
    'ResizeFcn',@regui, ...
    %  < ------ and so forth ------ >
     );
\end{lstlisting}

\begin{figure}[p]
\begin{center}
\includegraphics[width=0.7\textwidth]{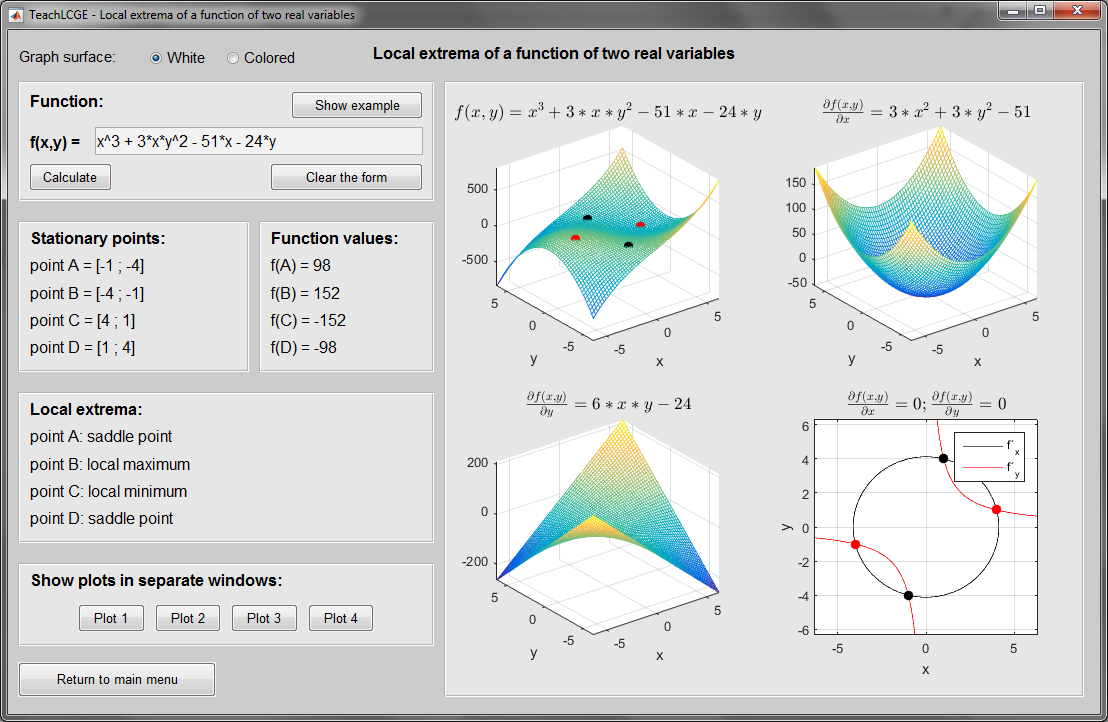}
\caption{Layout for the aspect ratio between 0.75 and 1.5.}
\label{fig:4to3}
\end{center}
\end{figure}
\begin{figure}[p]
\begin{center}
\includegraphics[width=0.7\textwidth]{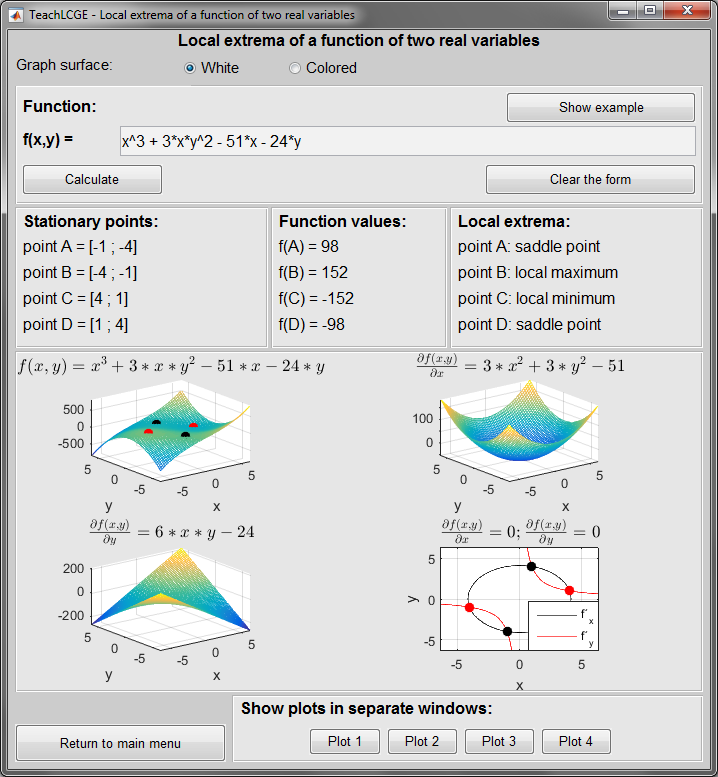}
\caption{Layout for the aspect ratio less than 0.75.}
\label{fig:vert-layout}
\end{center}
\end{figure}
\begin{figure}
\begin{center}
\includegraphics[width=0.85\textwidth]{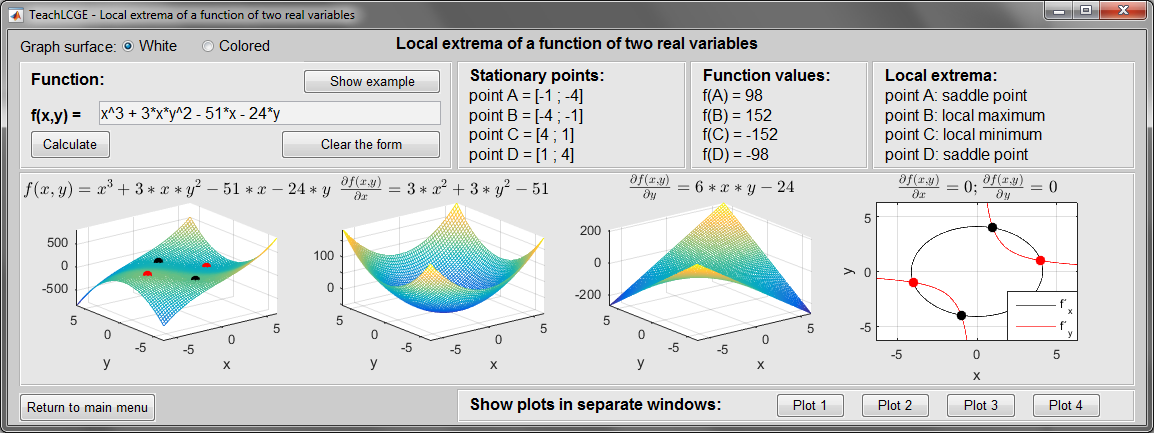}
\caption{Layout for the aspect ratio greater than 1.5.}
\label{fig:hor-layout}
\end{center}
\end{figure}

\newpage

\section{Conclusion}

In this paper we extended the idea of responsive design for the case of stand-alone applications, 
and illustrated this on a sample application created in MATLAB. The same technique can be 
used in other programming languages as well. 

Although in our example we -- for the sake of simplicity --  were changing only the layout 
of the logical blocks of the responsive graphical interface (ReGUI) of our sample application,
it is also possible to hide or show selected blocks of the ReGUI based on 
the current size of the application window or certain proportions of its width and height. 
Further, it is possible to change the color depth of the elements of the ReGUI, their visibility, 
their background (for example, switching between a solid color and a background image), 
and other visual and functional properties of the elements of ReGUI.
Also, in a similar manner the layout of the ReGUI of an application 
can change based on the position of the application on the computer screen -- 
for example, swap the left- and right-sided logical blocks
when a user moves the application window to the left or right edge of the screen.

\section*{Acknowledgments}

% Contribution of authors:
% Matej Mikulszky did the complete implementations in MATLAB 
% of all sample applications mentioned in this paper.
% Jana Pocsova and Andrea Mojzisova designed the didactical application.
% Igor Podlubny developed an ReGUI framework.

This work was  supported in parts by grants VEGA 1/0908/15,  
APVV-14-0892, SK-PL-2015-0038, ARO W911NF-15-1-0228.

.

\end{document}